# Programmable DNA-Mediated Decision Maker

**Jian-Jun SHU**[*], **Qi-Wen WANG and Kian-Yan YONG**

School of Mechanical & Aerospace Engineering, Nanyang Technological University, 50 Nanyang Avenue, Singapore 639798

ABSTRACT

DNA-mediated computing is a novel technology that seeks to capitalize on the enormous informational capacity of DNA and has tremendous computational ability to compete with the current silicon-mediated computing, due to massive parallelism and unique characteristics inherent in DNA interaction. In this paper, the methodology of DNA-mediated computing is utilized to enrich decision theory, by demonstrating how a novel programmable DNA-mediated normative decision-making apparatus is able to capture rational choice under uncertainty.

KEYWORDS: DNA; processor; material; programmable biochemical operator

INTRODUCTION

In nature, most of our daily activities, such as interacting with information received from the outside world and responding with appropriate reactions, are processed in our brains, which is a magnificent decision-making apparatus. The actual mechanism on how our genetic material, deoxyribonucleic acid (DNA), is translated into functional amino acids and eventually dedicated in the construction of sophisticated decision-making system still remains mysterious. The intrinsic characteristics of DNA, on the other hand, reveal the feasibility that artificially-synthesized DNA can be utilized as information storage and processing substrate, and ultimately assembled as a splendid normative decision-making apparatus, which is capable of making rational decision based on received information.

---

[*] Correspondence should be addressed to Jian-Jun SHU, mjjshu@ntu.edu.sg

DNA-mediated computational ability, through the in-vitro manipulation of artificially-encoded DNA sequences, has been successfully demonstrated to cope with some intriguing conundrums, including Hamiltonian path problem [1], satisfaction problem [2], maximal clique problem [3] and strategic assignment problem [4]. These ideas have ignited the potential power of assembling artificially-synthesized DNA as an astonishing computational device [5,6]. Most recently, further evidence reveals that DNA can be utilized in the construction of molecular logic circuits [7–10], simple artificial intelligent gaming device [11], and even the neural networks [12]. In this paper, a novel programmable DNA-mediated normative decision-making apparatus is proposed.

## Normative decision-making apparatus

Normative decision-making apparatus is a typical device, which is employed to simulate the rational behaves of individuals facing risky choices. Expected utility (EU) hypothesis [13], which is vastly used in the analysis of decision making under risk, is adopted as a normative model of rational choice.

The classical EU model of decision making under risk can be abstractly described as follows: Given a situation with a set of alternative options $O_i$ and by assuming that individual decision maker follows a series of predefined axioms, the rational choice is subjected to the option, whereas the EU $U$ as described by (1) is maximized.

$$U = \max \sum P_j u(x_j) \qquad (1)$$

where $P_j$ denotes the probability at each possible outcome $x_j$, and $u(x_j)$ denotes the utility of receiving outcome $x_j$.

The programmable DNA-mediated normative decision-making apparatus is employed to achieve the identical function by means of the in-vitro manipulation of artificially-encoded DNA molecules. A simple case study, which is modified from that of the Ellsberg paradox [14], is selected for demonstration purpose. It is a ball-selecting game, which involves 90 balls in an urn. The number of ball is known to the decision maker: 40 red balls (R), 30 black balls (B), and 20 white balls (W).

The decision maker is offered with three distinct gambling scenarios – *option 1*: receive $u(20)$ if a red or black ball is drawn and $u(0)$ if a white ball is drawn; *option 2*: receive $u(20)$ if a red or white ball is drawn and $u(0)$ if a black ball is drawn; *option 3*: receive $u(20)$ if a black or white ball is drawn and $u(0)$ if a red ball is drawn. In the remaining part of paper, $u(20)$ and $u(0)$ are interchangeably referred to as favorable outcome and unfavorable outcome, respectively. The question of the case study is which gambling scenario of three alternative options offers the best outcome to the decision maker. The entire case can be represented in terms of the decision matrix as shown in Table.

|  | Outcomes | | |
|---|---|---|---|
|  | $X_R=40$ | $X_B=30$ | $X_W=20$ |
| Probability | 4/9 | 1/3 | 2/9 |
| *Option 1* | $u(20)$ | $u(20)$ | $u(0)$ |
| *Option 2* | $u(20)$ | $u(0)$ | $u(20)$ |
| *Option 3* | $u(0)$ | $u(20)$ | $u(20)$ |

**Table**: *Decision matrix*

For the sake of convenience, the decision matrix can be transformed into an equivalent decision tree. The choice node stands for the point that decision maker picks up a gambling scenario from a set of predefined alternative options. The chance node is associated with various objectives probabilities leading to distinct outcomes. The chance node can be classified into two types, known as the favorable outcome $u(20)$, and unfavorable outcome $u(0)$. In addition to the decision tree, one additional node, which is named as termination node, is introduced in the graph. By doing so, the original decision tree is converted into a directed network, which begins at the choice node and ends at the termination node, as shown in Fig. 1.

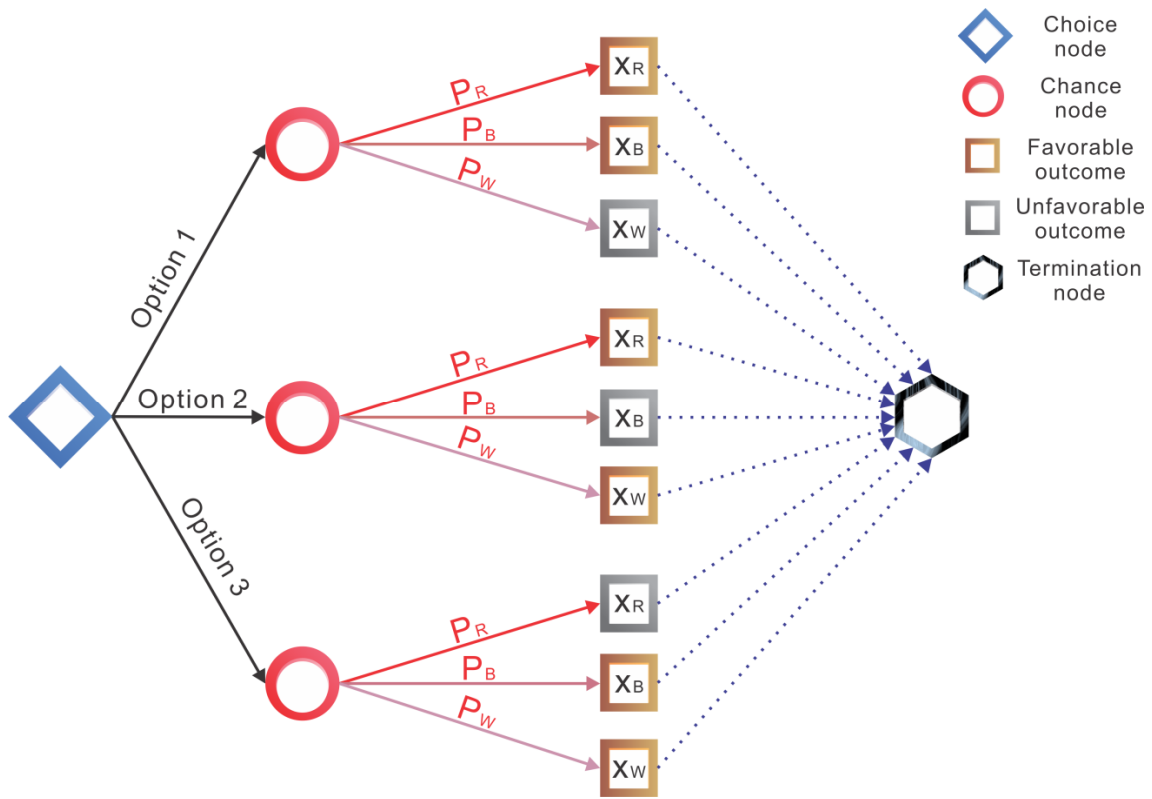

**Fig. 1**: *Decision tree*

PROBLEM ENCODING

As being analogous to the very beginning step of contemporary silicon-mediated computer, whereas information is digitized in terms of binary expressions, the information as depicted in Fig. 1 is converted into a combination of DNA sequences – Adenine (A), Thymine (T), Guanine (G), and Cytosine (C) – based upon the DNA sequence design motif specified in Fig. 2. As the several section of proposed design motif is similar as described in [4,5], a detailed explanation of similar parts is therefore eliminated in this paper. By doing so, it is possible to concentrate on the newly-created problem encoding strategies.

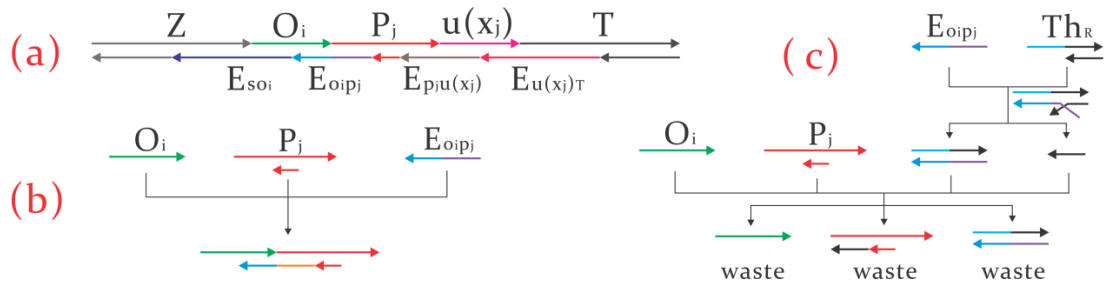

**Fig. 2**: *DNA sequence design motif*

It is to begin with the encoding scheme for option edges, as instructed in Fig. 1. Each edge of option $O_i$, where $i \in \{1,2,3\}$, is encoded with 20-mer single-stranded DNA (ssDNA). To avoid undesired hybridization in the subsequent procedure, all DNA sequences used to represent the option edges, and the remaining DNA sequences engaged in hybridization, are accomplished based on the rules as specified in [15]. In addition to that, each DNA sequence of $O_i$ contains a recognition site (as underlined in Fig. 2) with respect to one specific restriction enzyme – *option 1*: *Proteus vulgaris* (PvuII); *option 2*: *Haemophilus parainfluenzae* (HpaI); *option 3*: *Streptomyces tubercidicus* (StuI). The selected restriction enzymes have two properties in common – all these chosen enzymes are only effective on double-stranded DNA (dsDNA) and cleave original 20 base pair (bp) dsDNA into two dsDNA of 10 bp length ended up with blunt ends.

The probabilities of picking red ball ($P_R$), black ball ($P_B$), and white ball ($P_W$) as mentioned in decision tree (Fig. 1) are represented by introducing another class of DNA sequences, namely, threshold DNA sequence ($Th_j$), where $j \in \{R,B,W\}$. Without the presence of DNA sequence ($Th_j$), the DNA sequence of chance node ($E_{O_iP_j}$) is capable of joining the corresponding DNA sequence of option $O_i$ and probability $P_j$, as illustrated in Fig. 2 (b), due to its designated nature – the former 10-mer ssDNA is complementary to the rear 10-mer ssDNA of $O_i$, and the rear 10-mer ssDNA is complementary to the former 10-mer stick end of $P_j$. The obtained sequence promotes the formation of DNA sequence as described in Fig. 2 (a). By manually introducing the threshold DNA sequence ($Th_j$) into DNA solution, DNA sequence representing chance node ($E_{O_iP_j}$) is likely to react with the corresponding threshold DNA sequence ($Th_j$) due to toehold mediated strand displacement mechanism. The resultant DNA sequences containing option $O_i$ and probability $P_j$ lead to three distinct parts, as shown in Fig. 2 (c). All these parts are regarded as waste because none of them contributes to the formation of DNA sequence as described in Fig. 2 (a).

By following the toehold mediated strand displacement mechanism, the probabilities of picking red ball ($P_R$), black ball ($P_B$), and white ball ($P_W$), as instructed in Fig. 1, can be therefore represented by using various concentration of threshold DNA sequence ($Th_j$). The concentration ratio $R_i$ between threshold DNA sequence ($Th_j$) and chance node DNA sequence ($E_{O_i P_j}$) is estimated by using formula (2):

$$R_i = 1 - P_i \tag{2}$$

Therefore, in this study case, the ratios among threshold DNA sequences of red ball, black ball, and white ball are 5/9, 6/9, and 7/9, respectively.

As a result of gel electrophoresis, the DNA molecules of different lengths are used to indicate various concentrations. The DNA sequences representing probability ($P_j$), as instructed in Fig. 2, are the double-stranded DNA (dsDNA) with two identical-length sticky ends. The length of the middle section (dsDNA) is a variable, depending on the following rule: The concentration of DNA is inversely proportional to the assigned length of dsDNA. Based on the same ratio as discussed above, the length of the middle section of probability DNA sequences $P_j$ are set to be 7 bp for $P_R$, 16 bp for $P_B$, and 34 bp for $P_W$.

Finally, the outcome utility DNA sequence $u(x_j)$ is encoded with 20-mer single-stranded DNA (ssDNA). As being similar to the option DNA sequence $O_i$, each DNA sequence of outcome utility contains a recognition site (as underlined in Fig. 2) with respect to one specific restriction enzyme – $u(x_R)$: *Pseudomonas maltophilia* (PmlI); $u(x_B)$: *Escherichia coli* (EcoRV); $u(x_W)$: *Streptomyces caespitosus* (ScaI). The underlying principle is that, if the resultant utility of outcome is favorable, $u(x_j) = u(20)$, the presence of related restriction enzyme should be inhibited in the final mixture of DNA solutions. Otherwise, the assigned restriction enzyme is required to be added into the solution.

IMPLEMENTATION

After all DNA strands are synthesized in individual test tubes based on the DNA sequence design motif (Fig. 2), a small amount ($0.1 ml$ of solution at concentration of $0.1 \mu g / \mu l$) of DNA amount from all test tubes is extracted and mixed in one new test tube. At the end of hybridization, T4 DNA ligase is added into the solution, and sufficient time is allowed for the reaction.

The mixed solution is equally divided into three new test tubes, which are labelled as $TT1$ (test tube 1), $TT2$ (test tube 2), and $TT3$ (test tube 3). After that, three distinct restriction enzymes, HpaI, StuI, and ScaI, are added into $TT1$. The addition of the former two restriction enzymes, HpaI and StuI, are to cleave the DNA molecules, which contain the sequence representing the option $O_2$ and $O_3$. The concept is to retain only DNA molecules containing the specific sequence of option $O_1$. The addition of the last restriction enzyme ScaI is to eliminate the solution containing sequences representing the unfavorable outcome $u(0)$. By following the guideline, restriction enzymes, PvuII, StuI, EcoRV and PvuII, HpaI, PmlI are added into $TT2$ and $TT3$, respectively. All three test tubes, containing a mixture of DNA solutions, proper restriction enzyme, and appropriate master mix, are immersed into a water tank with water temperature maintaining constant at $37^o C$ for effective incubation. As a consequence of restriction enzyme digestion, the DNA sequences that do not satisfy the design paradigm are not amplified in the subsequent procedure, and ultimately, removed as the result of solution purification.

Polymerase chain reaction (PCR) is utilized to amplify, or to make the duplicate copies of target DNA sequence by varying the "input signal" known as the DNA primer. Two DNA oligonucleotides, the sequences " 3' CCTGGCTGTG 5' " and " 3' AGCGAGTGTT 5' ", are exploited as the DNA primers for all three test tubes. After $n$ complete thermal cycles of PCR, the amount of target DNA sequences is multiplied by $2^n$. PCR has two distinct features in this model: One is to selectively amplify and retain the DNA sequences beginning with choice node $Z$ and ending with termination node $T$; Another is to quantitatively enlarge concentration difference among amplified DNA sequences (prior to PCR, the initial concentration of amplified DNA solutions are distinct due to the involvement of threshold DNA sequences). For instance, the initial quantitative difference between any two DNA solutions is one unit. After, let us say, five complete thermal cycles of PCR, the difference is extended to 32 units, which is 32 times as compared with the initial difference. Such a difference can be easily captured by using spectrophotometer.

Prior to the gel electrophoresis, the solution obtained from PCR is subjected to the solution purifier, which significantly improves the separation outcome. The solution purifier is utilized to remove the "noise signal", including DNA waste due to toehold mediated strand displacement method, cleaved DNA sequences due to restriction enzymes, unamplified DNA sequences, excessive DNA oligonucleotide (primer), and other degradation factors like salts.

Purified DNA solution in all test tubes ($TT1$, $TT2$, and $TT3$) are inserted into a pre-made 4 – well agarose gel slab immersed in the TAE (a mixture of Tris base, acetic acid, and EDTA) buffer. For optimal DNA resolution, the recommended agarose gel concentration is 2.5 – 3%. The well labelled "1", "2", and "3" are used to hold the DNA solution with the same labelling number as that of test tubes. Well "4" contains the 10 bp DNA ladder, which holds 33 fragments ranging in size from 10 bp to 200 bp. After that, the gel slab is subjected to a constant electric field. Due to the working principle of gel electrophoresis, the final DNA sequence containing the shorter portion of probability sequence (higher probability) migrates faster towards anode electrode through the pores of gel matrix as compared with that of longer DNA probability sequence. For favorable resolution, the process terminates once the visible tracking dye (XCFF), whose migration speed is close to 100 bp double-stranded DNA, migrates 2/3 distance of the entire length of gel slab. Once the stained agarose gel is projected under the UV light of *302 nm* wavelength, each well (from well "1" to well "3") contains two DNA bands – well "1": 147 bp of relative concentration 4 and 156 bp of relative concentration 3; well "2": 147 bp of relative concentration 4 and 174 bp of relative concentration 2; well "3": 156 bp of relative concentration 3 and 174 bp of relative concentration 2.  Consequently, based on the observations obtained from the gel, it is possible to conclude that the gambling scenario 1 among all three alternatives offers the best possible outcome for the decision maker.

## CONCLUSIONS

DNA-mediated computing has inestimable potentials to be extended into other seemingly-unrelated disciplines, and ultimately changes the existing configurations in these fields.  In this paper, DNA-mediated computing has been correlated with the discipline of normative decision theory.  A case study is made to illustrate how to handle DNA-mediated decision making.  It is believed that the DNA-mediated technique proposed in this paper may facilitate effective complex decision-making in a rapidly changing environment.

## REFERENCES

1. Adleman, L.M., (1994). 'Molecular computation of solutions to combinatorial problems', *Science*, **266**(5187), 1021–1024.
2. Lipton, R.J., (1995). 'DNA solution of hard computational problems', *Science*, **268**(5210), 542–545.
3. Ouyang, Q., Kaplan, P.D., Liu, S.M. and Libchaber, A., (1997). 'DNA solution of the maximal clique problem', *Science*, **278**(5337), 446–449.
4. Shu, J.-J., Wang, Q.-W. and Yong, K.-Y., (2011). 'DNA-based computing of strategic assignment problems', *Physical Review Letters*, **106**(18), 188702.
5. Shu, J.-J., Wang, Q.-W., Yong, K.-Y., Shao, F. and Lee, K.J., (2015). 'Programmable DNA-mediated multitasking processor', *Journal of Physical Chemistry B*, **119**(17), 5639–5644.

6. Wong, J.R., Lee, K.J., Shu, J.-J. and Shao, F., (2015). 'Magnetic fields facilitate DNA-mediated charge transport', *Biochemistry*, **54**(21), 3392–3399.

7. Saghatelian, A., Völcker, N.H., Guckian, K.M., Lin, V.S.-Y. and Ghadiri, M.R., (2003). 'DNA-based photonic logic gates: AND, NAND, and INHIBIT', *Journal of the American Chemical Society*, **125**(2), 346–347.

8. Seelig, G., Soloveichik, D., Zhang, D.Y. and Winfree, E., (2006). 'Enzyme-free nucleic acid logic circuits', *Science*, **314**(5805), 1585–1588.

9. Frezza, B.M., Cockroft, S.L. and Ghadiri, M.R., (2007). 'Modular multi-level circuits from immobilized DNA-based logic gates', *Journal of the American Chemical Society*, **129**(48), 14875–14879.

10. Qian, L.L. and Winfree, E., (2011). 'Scaling up digital circuit computation with DNA strand displacement cascades', *Science*, **332**(6034), 1196–1201.

11. Pei, R.J., Matamoros, E., Liu, M.H., Stefanovic, D. and Stojanovic, M.N., (2010). 'Training a molecular automaton to play a game', *Nature Nanotechnology*, **5**(11), 773–777.

12. Qian, L.L., Winfree, E. and Bruck, J., (2011). 'Neural network computation with DNA strand displacement cascades', *Nature*, **475**(7356), 368–372.

13. Keeney, R.L. and Raiffa, H., (1993). Decisions with multiple objectives: Preferences and value tradeoffs, Cambridge University Press.

14. Ellsberg, D., (1961). 'Risk, ambiguity, and the savage axioms', *Quarterly Journal of Economics*, **75**(4), 643–669.

15. Tanaka, F., Kameda, A., Yamamoto, M. and Ohuchi, A., (2005). 'Design of nucleic acid sequences for DNA computing based on a thermodynamic approach', *Nucleic Acids Research*, **33**(3), 903–911.